\documentclass[twocolumn]{svjour3}          % twocolumn
\smartqed  % flush right qed marks, e.g. at end of proof
\usepackage{graphicx}
\usepackage[pdfborder={0 0 0},colorlinks=true,linkcolor=blue,citecolor=blue]{hyperref}

\newcommand{\sign}{\mathrm{sign}}

\begin{document}

\title{Spin-orbit coupling in Fe-based superconductors}

\author{M.M. Korshunov \and Yu.N. Togushova \and I. Eremin \and P.J. Hirschfeld}

%\authorrunning{M.M. Korshunov, Yu.N. Togushova, I. Eremin, P.J. Hirschfeld} % if too long for running head

\institute{M.M. Korshunov \at
              \email{korshunov@phys.ufl.edu} \\
              L.V. Kirensky Institute of Physics, Krasnoyarsk 660036, Russia
           \and
           M.M. Korshunov and Yu.N. Togushova \at
              Siberian Federal University, Svobodny Prospect 79, Krasnoyarsk 660041, Russia
           \and
           P.J. Hirschfeld \at
              Department of Physics, University of Florida, Gainesville, Florida 32611, USA
           \and
           I. Eremin \at
              Institut f\"ur Theoretische Physik III, Ruhr-Universit¨at Bochum, D-44801 Bochum, Germany \\
              Kazan Federal University, Kazan 420008, Russia
}

%\institute{M.M. Korshunov \at
%              Department of Physics, University of Florida, Gainesville, Florida 32611, USA \\
%              \email{korshunov@phys.ufl.edu} \\
%              \emph{Present addresses:} L.V. Kirensky Institute of Physics, Siberian Branch of RAS, 660036 Krasnoyarsk, and Siberian Federal University, 79 Svobodny Prospect, 660041 Krasnoyarsk, Russia
%           \and
%           Yu.N. Togushova \at
%              Siberian Federal University, 79 Svobodny Prospect, Krasnoyarsk 660041, Russia
%           \and
%           I. Eremin \at
%              Institut f\"ur Theoretische Physik III, Ruhr-Universit¨at Bochum, D-44801 Bochum, Germany
%           \and
%           P.J. Hirschfeld \at
%              Department of Physics, University of Florida, Gainesville, Florida 32611, USA
%}

\date{Received: date / Accepted: date}
% The correct dates will be entered by the editor

\maketitle

\begin{abstract}
We study the spin resonance peak in recently discovered iron-based
superconductors. The resonance peak observed in inelastic neutron scattering experiments agrees well with predicted results for the extended $s$-wave ($s_\pm$) gap symmetry. Recent neutron scattering measurements show that there is a disparity between longitudinal and transverse components of the dynamical spin susceptibility. Such breaking of the spin-rotational invariance in the spin-liquid phase can occur due to spin-orbit coupling. We study the role of the spin-orbit interaction in the multiorbital model for Fe-pnictides and show how it affects the spin resonance feature.
% \PACS{PACS code1 \and PACS code2 \and more}
% \subclass{MSC code1 \and MSC code2 \and more}
\end{abstract}

The nature of the superconductivity and gap symmetry and structure in the recently discovered Fe-based superconductors (FeBS) are the most debated topics in condensed matter community \cite{ROPPreview2011}. These quasi two-dimensional systems shows a maximal $T_c$ of 55\,K placing them right after high-$T_c$ cuprates. Fe $d$-orbitals form the Fermi surface (FS) which in the undoped systems consists of two hole and two electron sheets. Nesting between these two groups of sheets is the driving force for the spin-density wave (SDW) long-range magnetism in the undoped FeBS and the scattering with the wave vector $\mathbf{Q}$ connecting hole and electron pockets is the most probable candidate for superconducting pairing in the doped systems. In the spin-fluctuation studies \cite{Graser,Kuroki,Maiti}, the leading instability is the extended $s$-wave gap which changes sign between hole and electron sheets ($s_{\pm}$ state) \cite{Mazin_etal_splusminus}.

Neutron scattering is a powerful tool to measure dynamical spin susceptibility $\chi(\mathbf{q},\omega)$. It carries information about the order parameter symmetry and gap structure. For the local interactions (Hubbard and Hund's exchange), $\chi$ can be obtained in the RPA from the bare electron-hole bubble $\chi_0(\mathbf{q},\omega)$ by summing up a series of ladder diagrams to give
%
%\begin{eqnarray}
$\chi(\mathbf{q},\omega) = \left[I - U_s \chi_0(\mathbf{q},\omega)\right]^{-1} \chi_0(\mathbf{q},\omega)$,
%\label{eq:chi_s_sol}
%\end{eqnarray}
%
where $U_s$ and $I$ are interaction and unit matrices in orbital space, and all other quantities are matrices as well.

\begin{figure}[ht]
\begin{center}
\includegraphics[width=.48\textwidth]{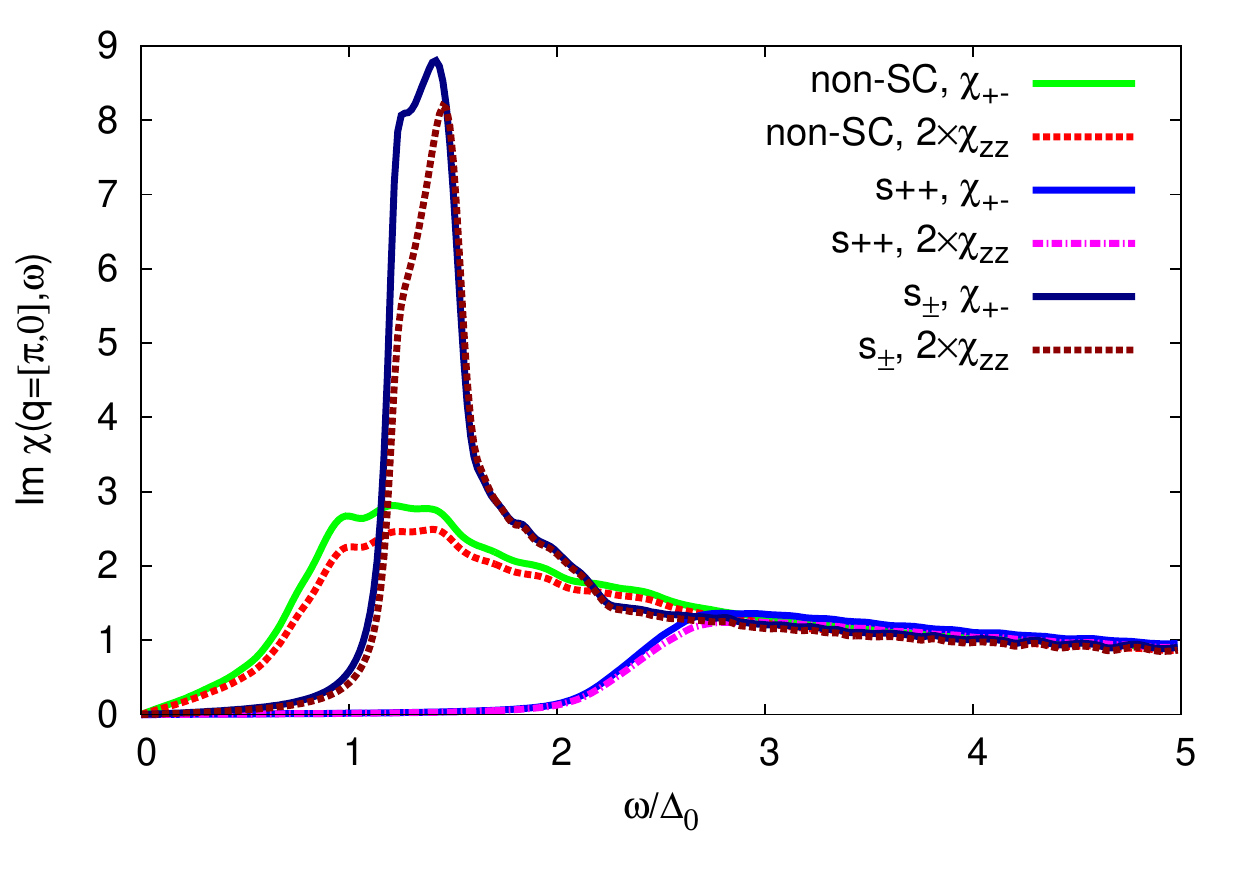}
\caption{Fig.~1. Calculated $\mathrm{Im}\chi(\mathbf{Q},\omega)$ in the normal state, and for the $s_{++}$ and $s_\pm$ pairing symmetries. In the latter case, the resonance is clearly seen below $\omega = 2\Delta_0$. Spin-orbit coupling constant $\lambda=100$\,meV, intraorbital Hubbard $U=0.9$\,eV, Hund's $J=0.1$\,eV, interorbital $U'=U-2J$, and pair-hopping term $J'=J$.
\label{fig:F1}}
\end{center}
\end{figure}

Scattering between nearly nested hole and electron Fermi surfaces in FeBS produce a peak in the normal state magnetic susceptibility at or near $\mathbf{q} = \mathbf{Q}=(\pi,0)$. For the uniform $s$-wave gap, $\sign \Delta_\mathbf{k} = \sign \Delta_{\mathbf{k}+\mathbf{Q}}$ and there is no resonance peak. For the $s_\pm$ order parameter as well as for an extended non-uniform $s$-wave symmetry, $\mathbf{Q}$ connects Fermi sheets with the different signs of gaps. This fulfills the resonance condition for the interband susceptibility, and the spin resonance peak is formed at a frequency below $\Omega_c = \min \left(|\Delta_\mathbf{k}| + |\Delta_{\mathbf{k}+\mathbf{q}}| \right)$ (compare normal and $s_\pm$ superconductor's response in Fig.~\ref{fig:F1}) \cite{Korshunov2008,Maier,Maier2}. The existence of the spin resonance in FeBS was predicted theoretically \cite{Korshunov2008,Maier} and subsequently discovered experimentally with many reports of well-defined spin resonances in 1111, 122, and 11 systems \cite{ChristiansonBKFA,Inosov,Argyriou}.

One of the recent puzzles in FeBS is the discovered anisotropy of the spin resonance peak in Ni-doped Ba-122 \cite{Lipscombe2010}. It was found that $\chi_{+-}$ and  $2\chi_{zz}$ are different. This contradicts the spin-rotational invariance (SRI) $\left<S_+ S_-\right> = 2\left<S_z S_z\right>$ which have to be obeyed in the disordered system. One of the solution to the puzzle is the spin-orbit (SO) interaction which can break the SRI like it does in Sr$_2$RuO$_4$ \cite{Eremin2002}. Here we incorporate the effect of the SO coupling in the susceptibility calculation for FeBS to shed light on the spin resonance anisotropy.

The simplest model for pnictides in the 1-Fe per unit cell Brillouin zone comes from the three $t_{2g}$ $d$-orbitals. The $xz$ and $yz$ components are hybridized and form two electron-like FS pockets around $(\pi,0)$ and $(0,\pi)$ points, and one hole-like pocket around $\Gamma=(0,0)$ point. The $xy$ orbital is considered to be decoupled from them and form an outer hole pocket around $\Gamma$ point. The one-electron part of the Hamiltonian is given by
%
%\begin{equation}\label{H0}
$H_0=\sum\limits_{\mathbf{k},\sigma,l,m} \varepsilon_\mathbf{k}^{l m} c_{\mathbf{k} l \sigma}^\dag c_{\mathbf{k} m \sigma}$,
%\end{equation}
%
where $l$ and $m$ are orbital indices, $c_{\mathbf{k} m \sigma}$ is the annihilation operator of a particle with momentum $\mathbf{k}$ and spin $\sigma$.
%
%Matrix of the one-electron energies is
%%
%\begin{equation}\label{epsM3orb}
% \hat\varepsilon_\mathbf{k} = \left(
%                    \begin{array}{ccc}
%                      \varepsilon_{1\mathbf{k}} & 0 & 0 \\
%                      0 & \varepsilon_{2\mathbf{k}} & \varepsilon_{4\mathbf{k}} \\
%                      0 & \varepsilon_{4\mathbf{k}} & \varepsilon_{3\mathbf{k}} \\
%                    \end{array}
%                  \right).
%\end{equation}
%
%Here
%%
%\begin{eqnarray}\label{eps3orb}
% \varepsilon_{1\mathbf{k}} &=& \epsilon_{xy} - \mu + 2 t_{xy} (\cos{k_x}+\cos{k_y}) + 4 t_{xy}' \cos{k_x}\cos{k_y}, \\
% \varepsilon_{2\mathbf{k}} &=& \epsilon_{yz} - \mu + 2 t_x \cos{k_x} + 2 t_y \cos{k_y} + 4 t' \cos{k_x}\cos{k_y} + 2 t'' (\cos{2 k_x}+\cos{2 k_y}), \\
% \varepsilon_{3\mathbf{k}} &=& \epsilon_{xz} - \mu + 2 t_y \cos{k_x} + 2 t_x \cos{k_y} + 4 t' \cos{k_x}\cos{k_y} + 2 t'' (\cos{2 k_x}+\cos{2 k_y}), \\
% \varepsilon_{4\mathbf{k}} &=& 4 t_{xzyz} \sin{k_x/2}\sin{k_y/2}.
%\end{eqnarray}
%%
%To reproduce the topology of the FS in pnictides, we choose the following parameters (in eV): $\mu=0, \epsilon_{xy}=-0.70, \epsilon_{yz}=-0.34, \epsilon_{xz}=-0.34, t_{xy}=0.18, t_{xy}'=0.06, t_x=0.26, t_y=-0.22, t'=0.2, t''=-0.07, t_{xzyz}=0.38$.
%
This model for pnictides is similar to the one for Sr$_2$RuO$_4$ and, in particular, the $xy$ band does not hybridize with the $xz$ and $yz$ bands. Keeping in mind the similarity to the Sr$_2$RuO$_4$ case, for simplicity we consider only the $L_z$-component of the SO interaction \cite{Eremin2002}. Due to the structure of the
%SO coupling's
$L_z$-component, the interaction affects $xz$ and $yz$ bands only.

Following Ref.~\cite{Ng2000}, we write the SO coupling term, $H_{SO} = \lambda \sum\limits_{f} \vec{L}_f \cdot \vec{S}_f$, in the second-quantized form as
%
%\begin{equation}\label{HSO}
 $H_{SO} = \mathrm{i} \frac{\lambda}{2} \sum\limits_{l,m,n} \epsilon_{l m n} \sum\limits_{\mathbf{k}, \sigma, \sigma'} c_{\mathbf{k} l \sigma}^\dag c_{\mathbf{k} m \sigma'} \hat{\sigma}_{\sigma\sigma'}^n$,
%\end{equation}
%
where $\epsilon_{l m n}$ is the completely antisymmetric tensor, indices $\{l,m,n\}$ take values $\{x,y,z\} \leftrightarrow \{d_{yz},d_{zx},d_{xy}\} \leftrightarrow \{2,3,1\}$,  and $\hat{\sigma}_{\sigma\sigma'}^n$ are the Pauli spin matrices.

The matrix of the Hamiltonian $H = H_0+H_{SO}$ is then
\begin{equation}\label{epsks_z}
 \hat\varepsilon_{\mathbf{k} \sigma} = \left(
                      \begin{array}{ccc}
                        \varepsilon_{1\mathbf{k}} & 0 & 0 \\
                        0 & \varepsilon_{2\mathbf{k}} & \varepsilon_{4\mathbf{k}} + \mathrm{i} \frac{\lambda}{2} \mathrm{sign}{\sigma} \\
                        0 & \varepsilon_{4\mathbf{k}} - \mathrm{i} \frac{\lambda}{2} \mathrm{sign}{\sigma} & \varepsilon_{3\mathbf{k}} \\
                      \end{array}
                     \right)
\end{equation}
%
%Corresponding eigenvalues are
%%
%\begin{eqnarray}\label{E_z}
% E_{1\mathbf{k}} &=& \varepsilon_{1\mathbf{k}},\\
% E_{2,3\mathbf{k}} &=& \frac{\varepsilon_{2\mathbf{k}}+\varepsilon_{3\mathbf{k}}}{2} \pm \sqrt{ \left(\frac{\varepsilon_{2\mathbf{k}}-\varepsilon_{3\mathbf{k}}}{2}\right)^2 + \varepsilon_{4\mathbf{k}}^2 + \frac{\lambda^2}{4}}.
%\end{eqnarray}
%%
As for Sr$_2$RuO$_4$, eigenvalues of $\hat\varepsilon_{\mathbf{k} \sigma}$ do not depend on spin $\sigma$, therefore, spin-up and spin-down states are still degenerate in spite of the SO interaction.

We calculated both $+-$ (longitudinal) and $zz$ (transverse) components of the spin susceptibility and found that in the normal state $\chi_{+-} > 2\chi_{zz}$ at small frequencies, see Fig.~\ref{fig:F1}. As expected, for the $s_{++}$ superconductor (conventional isotropic $s$-wave) there is no resonance peak and the disparity between $\chi_{+-}$ and $2\chi_{zz}$ is very small. For the $s_\pm$ superconductor, however, the situation is opposite -- we observe a well defined spin resonance and $\chi_{+-}$ is larger than $2\chi_{zz}$ by about 15\% near the peak position (Fig.~\ref{fig:F1}).

\textit{In summary}, we have shown that the spin resonance peak in FeBS gains anisotropy in the spin space due to the spin-orbit coupling. This result is in qualitative agreement with experimental findings. We do not observe changes in the peak position but this may be due to the simple model that we studied.

%Components of Eq.~(\ref{eq.chi1}) becomes
%%
%\begin{eqnarray}\label{eq.chizz}
% \chi^{ll',mm'}_{zz}(q,\mathrm{i}\Omega) &=& -\frac{T}{4} \sum_{\omega_n,p,\sigma} \left[ G_{m l \sigma}(p,\mathrm{i}\omega_n) G_{l' m' \sigma}(p+q,\mathrm{i}\Omega+\mathrm{i}\omega_n) \right.\nonumber\\
% &&\left. +F^\dag_{l m' \sigma}(p,-\mathrm{i}\omega_n) F_{l' m \sigma}(p+q,\mathrm{i}\Omega+\mathrm{i}\omega_n) \right].
%\end{eqnarray}
%%
%%
%\begin{eqnarray}\label{eq.chipm}
% \chi^{ll',mm'}_{+-}(q,\mathrm{i}\Omega) &=& -T \sum_{\omega_n,p,\sigma} \left[ G_{m l \uparrow}(p,\mathrm{i}\omega_n) G_{l' m' \downarrow}(p+q,\mathrm{i}\Omega+\mathrm{i}\omega_n) \right.\nonumber\\
% &&\left. -F^\dag_{l m' \uparrow}(p,-\mathrm{i}\omega_n) F_{l' m \downarrow}(p+q,\mathrm{i}\Omega+\mathrm{i}\omega_n) \right].
%\end{eqnarray}
%%

\begin{acknowledgements}
Partial support was provided by DOE DE-FG02-05ER46236 (P.J.H. and M.M.K.) and NSF-DMR-1005625 (P.J.H.). M.M.K. acknowledge support from RFBR (grants 09-02-00127, 12-02-31534 and 13-02-01395), Presidium of RAS program ``Quantum mesoscopical and disordered structures'' N20.7, FCP Scientific and Research-and-Educational Personnel of Innovative Russia for 2009-2013 (GK 16.740.12.0731 and GK P891), and President of Russia (grant MK-1683.2010.2), Siberian Federal University (Theme N F-11), Program of SB RAS \#44, and The Dynasty Foundation and ICFPM. I.E. acknowledges support of the SFB Transregio 12, Merkur Foundation, and German Academic Exchange Service (DAAD PPP USA No. 50750339).
\end{acknowledgements}

\end{document}